\documentstyle[floats,twocolumn,aps,prl,graphicx]{revtex}

\def\figun{
  \begin{figure}
    \begin{center}
      \includegraphics[width=8.5cm]{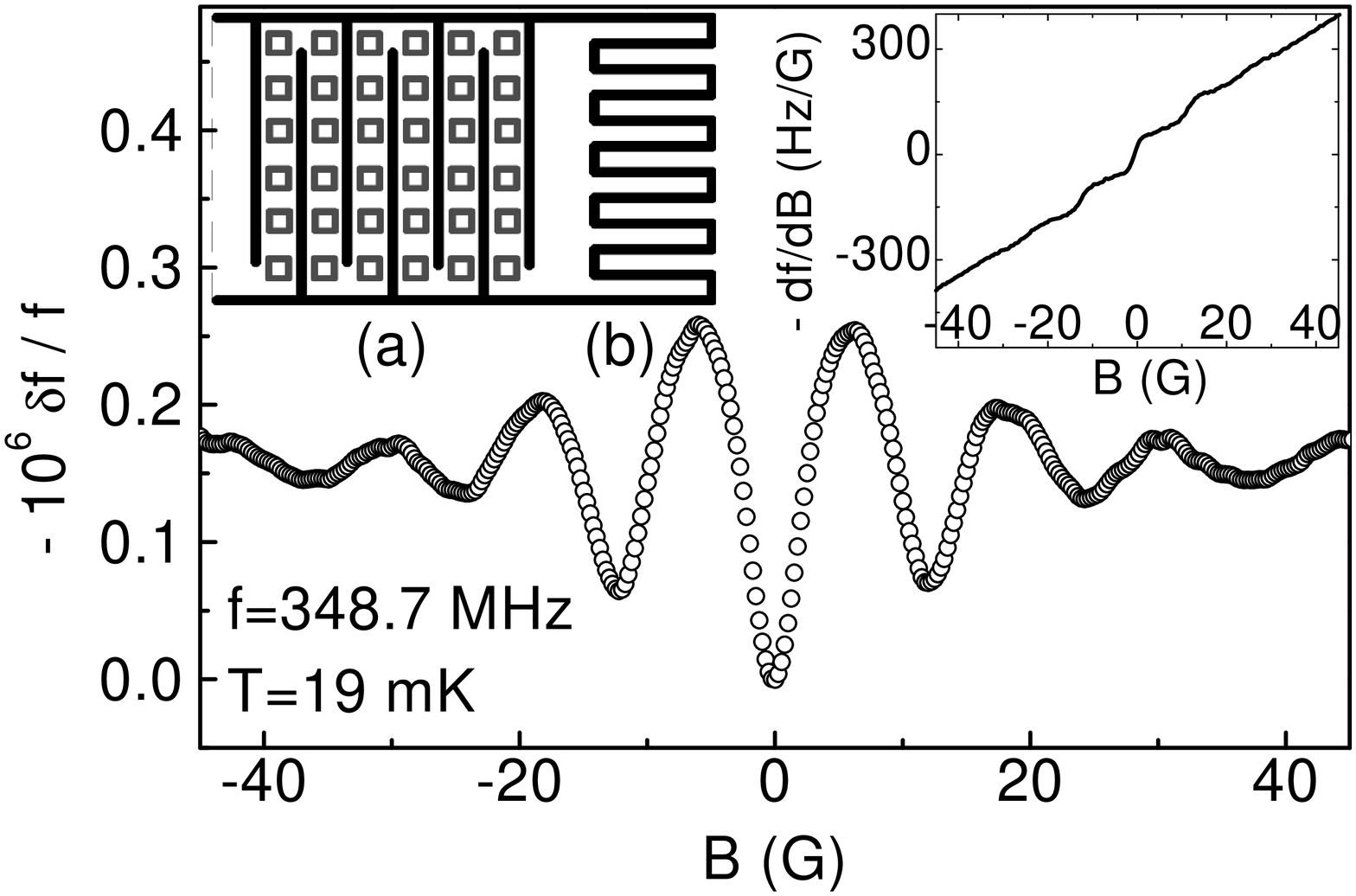}
      \caption{Typical relative variation of the resonance frequency
        (proportional to $\delta_\Phi \alpha^{'}$) versus magnetic
        field (Illumination 3). $\delta f/f$ is periodic with a period
        of 12.5 Gauss corresponding to half a flux quantum in a
        ring. Left Inset : schematic picture of the resonator and the
        rings on the capacitance of the resonator. The capacitance
        (comb like structure (a)) is well separated from the
        inductance (meander line (b)). Right Inset :
        derivative of the resonance frequency versus magnetic field
        before substraction of base line.}
      \label{fig:df}
    \end{center} 
  \end{figure}
}

\def\figdeux{
  \begin{figure}
    \begin{center}
      \includegraphics[width=8.5cm]{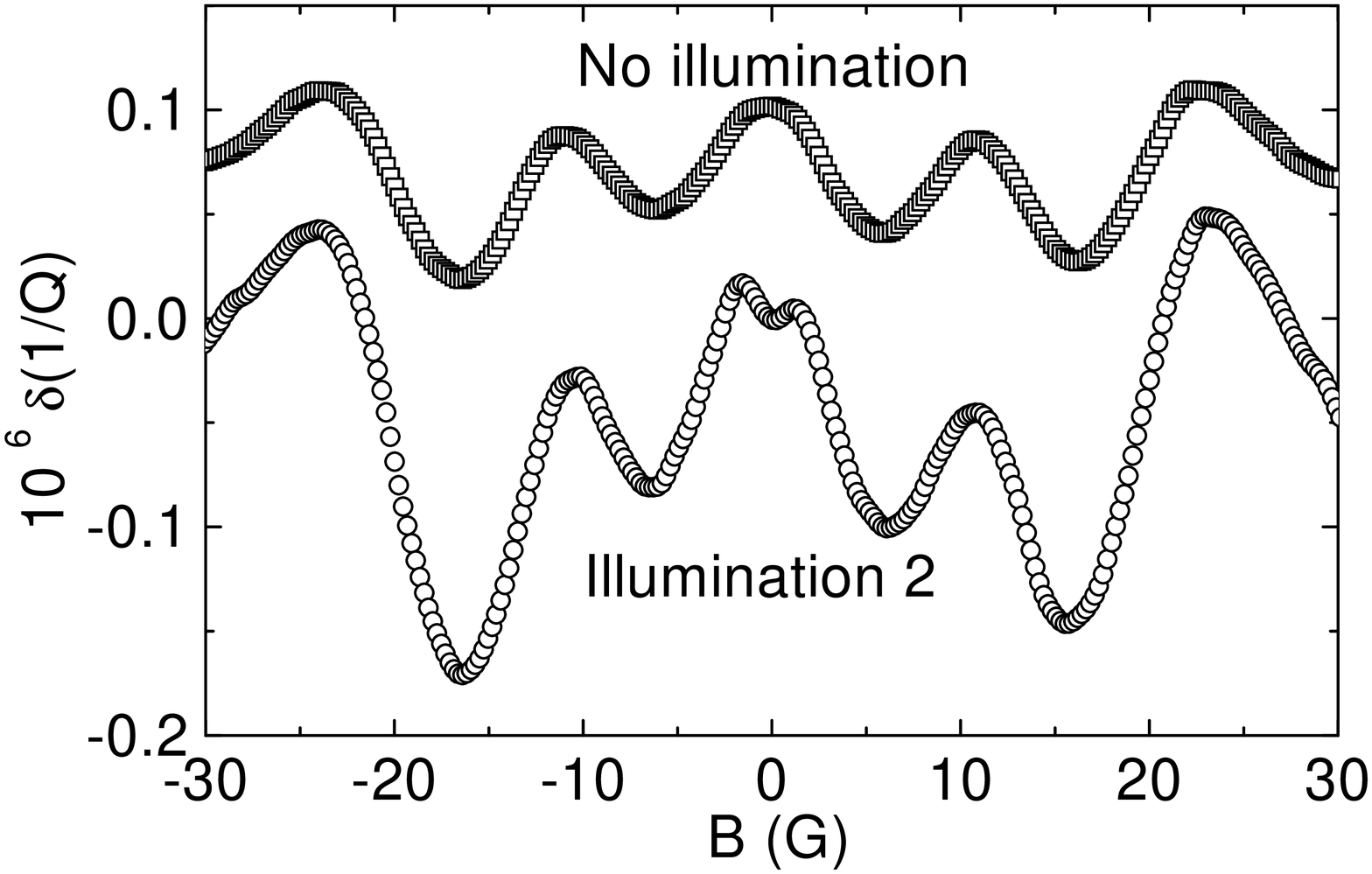}
      \caption{Variation of $1/Q$
        (proportional to $\delta_\Phi \alpha^{''}$) versus B at two
        different illuminations. The period of the oscillations is
        12.5 Gauss. At high illumination a dip appears in the zero
        field region. The curve at illumination 0 is shifted for
        clarity.}
      \label{fig:dq}
    \end{center}
  \end{figure}
}

\def\figtrois{
  \begin{figure}
    \begin{center}
      \includegraphics[width=8.5cm]{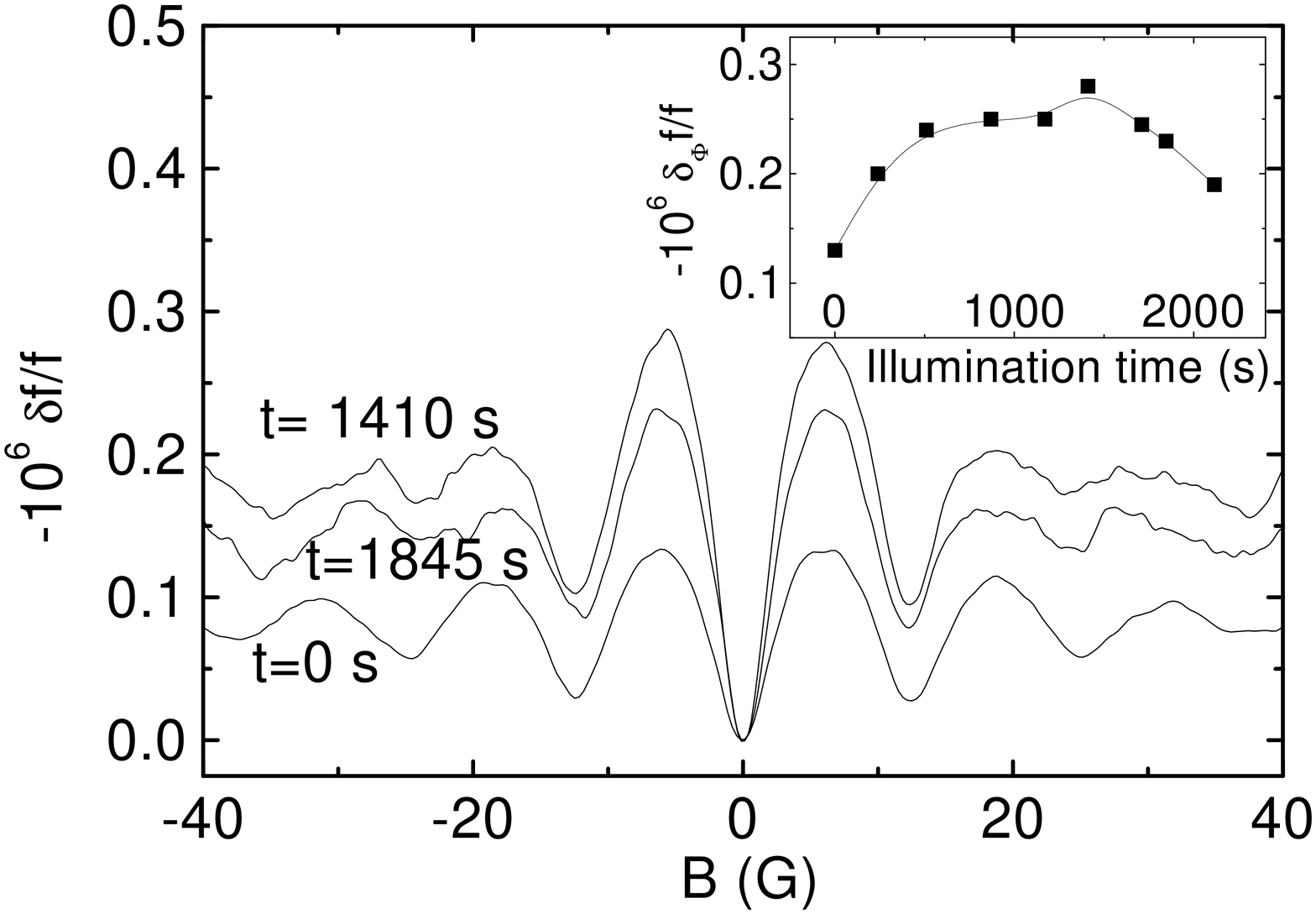}
      \caption{$\delta f/f$ versus B at different
        illuminations. Inset : Relative variation of resonance
frequency versus illumination time (continuous line is only a guide to
the eye).}
      \label{fig:dfillum}
    \end{center}
  \end{figure}
}

\def\figquatre{
  \begin{figure}
    \begin{center}
      \includegraphics[width=8.5cm]{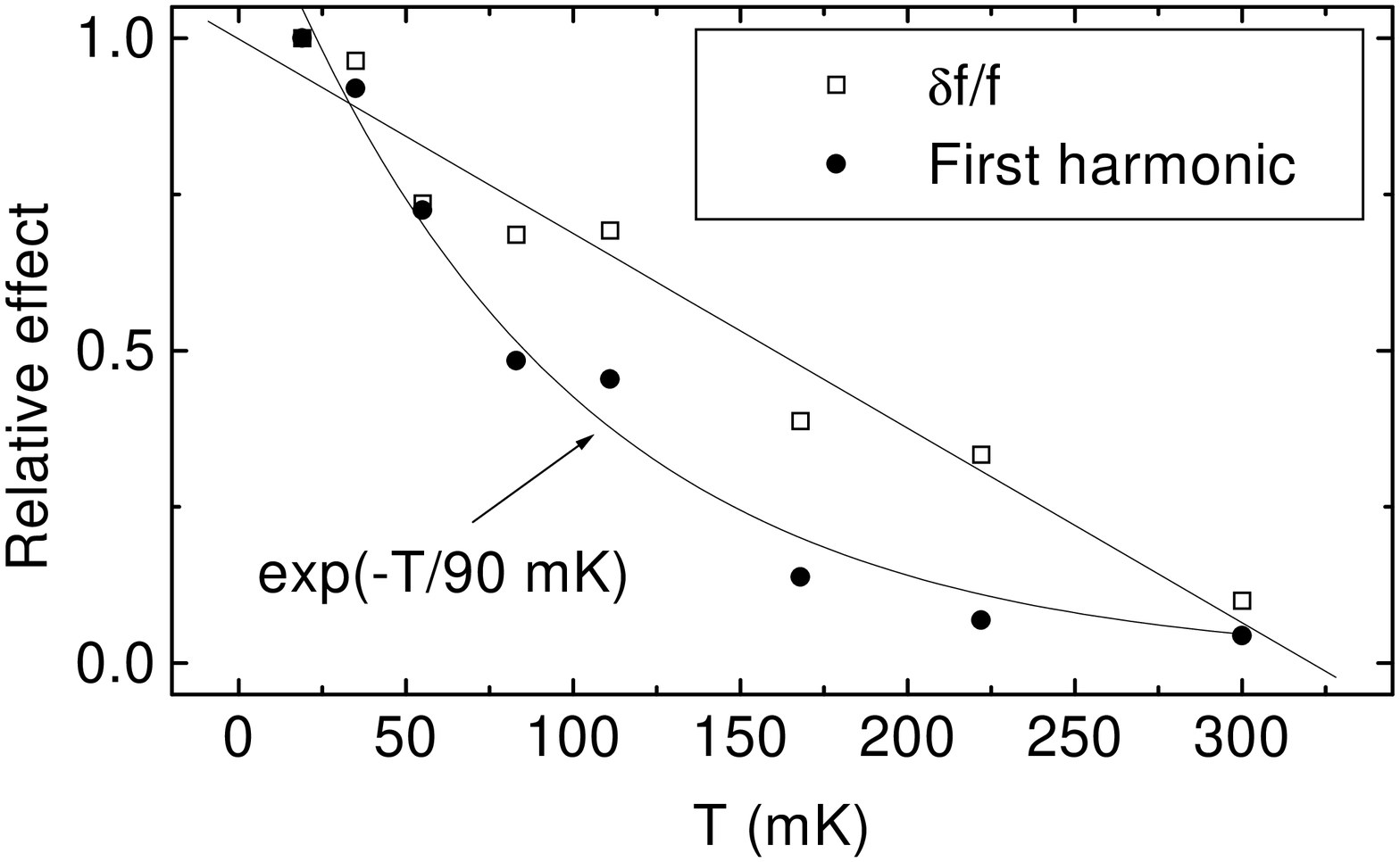}
      \caption{Temperature dependence of the relative variation of the
        resonance frequency and its first harmonic for the
        illumination 6. $\delta_\Phi f/f$ decreases linearly with
        temperature, whereas the first harmonic can be fitted by an
        exponential decay with a temperature scale of 90 mK,
        independent of illumination.}
      \label{fig:dftemp}
    \end{center}
  \end{figure}
}

\title{Measurements of flux dependent screening in Aharonov-Bohm
rings}

\author{R. Deblock$^1$, Y. Noat$^1$, H. Bouchiat$^1$, B. Reulet$^1$
  and D. Mailly$^2$}

\address{$^1$Laboratoire de Physique des Solides, Associ\'e au CNRS,
  B\^atiment 510, Universit\'e Paris-Sud, 91405, Orsay, France.\\
  $^2$CNRS Laboratoire de Microstructure et Micro\'electronique, 196
  avenue Ravera, 92220, Bagneux, France.\\
\parbox{14cm}{\medskip\rm\small
In order to investigate the effect of electronic phase coherence on
screening we have measured the flux dependent polarizability of
isolated mesoscopic rings at 350 MHz. At low temperature (below 100
mK) both non-dissipative and dissipative parts of the polarizability
exhibit flux oscillations with a period of half a flux quantum in a
ring. The sign and amplitude of the effect are in good agreement with
recent theoretical predictions. The observed positive
magneto-polarizability corresponds to an enhancement of screening when
time reversal symmetry is broken. The effect of electronic density and
temperature are also measured.}}
\begin{document}

\maketitle

When an electric field {\bf E} is applied to an isolated metallic
sample, electron screening gives rise to an induced dipole {\bf d}. In
the linear response regime :
\begin{equation}
{\bf d}=\alpha {\bf E}
\end{equation}
where $\alpha$ is the electric polarizability.  For a sample of
typical size $a$ much larger than the Thomas-Fermi screening length
$\lambda_s$, $\alpha$ is essentially determined by geometry, with a
negative correction of the order of $\lambda_s/a$ \cite{Rice}.  The
measurement of $\alpha$ gives information on the way electrons screen
an external electric field.  At the mesoscopic scale, when phase
coherence through the sample is achieved (i.e. the phase coherence
length is of the same order as the typical size of the system)
electronic properties are sensitive to the phase of the electronic
wave functions, which can be tuned by an Aharonov-Bohm flux in a ring
geometry \cite{Intro}. It has been recently suggested that screening
of an electric field may be sensitive to this phase coherence, leading
to a flux dependent mesoscopic correction to polarizability
\cite{Efetov,NRB,Blanter}. In particular the polarizability of an
Aharonov-Bohm ring is expected to exhibit oscillations as a function
of the magnetic flux.

We have inferred the electrical response of an array of rings from the
flux dependence of the capacitance $C$, placed underneath the rings,
of an RF superconducting micro-resonator(Fig \ref{fig:df}, left
inset). This experiment has been checked to be only sensitive to the
electrical response of the rings~ \cite{testRPE}. $C$ is modified by
the non dissipative response $\alpha^{'}$ of the
rings~{\cite{deblock}:  
\begin{equation} \label{dC}
\frac{\delta C}{C} = k N_s \alpha^{'}
\end{equation}
where $N_s$ is the number of rings coupled to the resonator and $k$
the electric coupling coefficient between a ring and the capacitance,
which only depends on geometry. Since the resonance frequency $f=1/(2
\pi \sqrt{L C})$, with $L$ the inductance of the resonator, the change
in $C$ shifts the resonance frequency. In addition the dissipative
response $\alpha^{''}$ of the rings weakens the quality factor
$Q$~:
\begin{equation} \label{dfdQ}
  \frac{\delta f}{f}=- \frac{1}{2} k N_{s} {\alpha}^{'}(\omega),
  \delta(\frac{1}{Q}) = k N_{s} {\alpha}^{''} (\omega)
\end{equation}
}
In order to measure $\delta f$ and $\delta Q$, the RF frequency is
modulated at 100 kHz. Using lock-in detection, the reflected signal
from the resonator at the modulated RF frequency is measured and used
to lock the experimental setup on the resonance frequency.  The
feedback signal is then proportional to the variation of $f$, whereas
the signal at double the modulation frequency is proportional to the
variation of $Q$.  We are careful to inject sufficiently low power
($\approx 10pW$) so as not to heat the sample.  The signal is measured
as a function of magnetic field.  To improve accuracy, the
derivative of the signal is also detected by modulating at 30 Hz a
magnetic field of 1 Gauss amplitude.  Our precision with this setup is
$\delta f/f=10^{-8}$ and $\delta Q/Q=10^{-8}$.

The rings are etched in a high-mobility GaAs-AlGaAs
heterojunction. Etching strongly decreases the conductivity of the
rings, but the nominal conductivity is recovered by illuminating the
sample with an infrared diode \cite{Diode}. We have checked this on a
connected sample. The characteristics of the rings, deduced from
transport measurements on wires of the same width etched in the same
heterojunction, are the following~: at nominal electronic density
($n_e=3.10^{11} $cm$^{-2}$) the mean free path $l_e=3 \mu$m, the
etched width is 0.5 $\mu$m whereas the effective width $W=$0.2 $\mu$m
(estimated from weak localization experiments \cite{EPL}) is much
smaller due to depletion, the coherence length is $L_{\Phi}=5.5 \mu$m
and the effective perimeter $L=$5.2 $\mu$m.  The rings are thus
ballistic in the transverse direction and diffusive
longitudinally. The mean level spacing $\Delta=h^2/(2 \pi m W L)
\simeq 80$ mK $\simeq 1.66$ GHz and the Thouless energy $E_c=h
D/L^2=450$ mK, with $D$ the diffusion coefficient and $m$ the
effective mass of electrons. The Thomas-Fermi screening length is
$\lambda_s=\pi a^{\star}/2=16$~nm, with $a^\star$ the effective Bohr
radius. There are $N=10^5$ rings on one sample. The resonator is made
by optical lithography in niobium on a sapphire substrate. The length
of the capacitance is $l=20.5$ cm and the inductance is 5 cm long. It
has a single resonance frequency $f_0 = 385$ MHz and a quality factor
of 10 000. The distance between the capacitance and the inductance is
300 $\mu$m. A 0.9 $\mu$m thick mylar film is inserted between the
detector and the ring substrate in order to reduce inhomogeneities
of the magnetic field due to the Meissner effect in the vicinity of
the superconducting resonator. The system constituted by the resonator
and the sample has a reduced quality factor of 3000, probably to dielectric
losses in the GaAs substrate, and a resonance frequency of 350
MHz. The system is cooled in a dilution refrigerator down to 18 mK.
\figun

The typical field dependence of the rings contribution to $\delta f/f$
is shown on figure \ref{fig:df}. This signal is superimposed on the
diamagnetic response of the niobium resonator (Fig \ref{fig:df},
right inset) which we substract in the following way : the base line
of the derivative of the resonance frequency is removed and the signal
is then integrated. One thus obtains the curves of figures 1 and 2,
which are directly proportional to the flux dependence of the
polarizability of the rings. The resonance frequency is periodic in
field with a period of 12.5 G which corresponds to half a flux quantum
$\Phi_0/2=h/2e$ in a ring, consistent with an Aharonov-Bohm effect
averaged over many rings \cite{Phi}. The resonance frequency decreases
by about 100Hz between 0 and 6 G. The magnetic field reduces the
amplitude of the oscillations, which cannot be detected for a field
higher than 40 G, consistent with the finite width of the ring.
Illuminating the sample with the electroluminescent diode strongly
affects the measured signal.  At low illumination the amplitude of the
oscillations increases and decreases at higher illuminations (see Fig
\ref{fig:dfillum}). The extremum amplitude of $\delta_\Phi f/f =
(f(\Phi=\Phi_0/4)-f(\Phi=0))/f$, is $-2.8 \, 10^{-7}$. In our geometry
one has \cite{deblock}:
\begin{equation} \label{k}
k = \frac{1}{\pi \epsilon_0 \epsilon a l d}
\frac{\ln(\frac{d+a}{d-a})}{\ln(\frac{d}{r})}
\end{equation}
with the relative dielectric constant of GaAs $\epsilon=12.85$, the
size of the rings $a=1.3\,\mu$m, the length of the capacitance $l=20.5$
cm, the distance between one lead of the capacitance and a ring $d/2
\simeq 3.15\, \mu$m and the width of the lead $r=1\, \mu$m. Since all the
rings are not identically coupled to the resonator, $k$ has to be
understood as an average capacitive coupling between one ring an the
capacitance. Note that approximatively half the rings are well
coupled to the capacitance. Because of these approximations the
experimental value is given with a $50 \%$ error range.  One obtains
$\delta_\Phi \alpha'/\alpha_{1D}=0.7 \, 10^{-3}$, where
$\alpha_{1D}=\epsilon_0 \pi^2 R^3/\ln(R/W)$ is the polarizability of a
quasi one dimensional (quasi-1D) circular ring of radius R.
\figdeux

The dissipative part of the polarizability is obtained from the field
dependence of $Q$ at different illuminations (Fig \ref{fig:dq}). It
exhibits a periodic behavior with the same period as the resonance
frequency. At low illumination $1/Q$ decreases with magnetic field for
small field whereas at higher illumination a dip in the zero field
region, which is not understood, appears indicating an increase of
$1/Q$ with field.  The typical amplitude of $\delta_\Phi(1/Q)$ is
$-10^{-7}$, hence $\delta_\Phi \alpha^{''}/\alpha_{1D}=-1.3 \,
10^{-4}$.

The sensitivity of the electrostatic properties of mesoscopic systems
to quantum coherence has been emphasized by B\"uttiker for connected
geometries \cite{Buttiker}. The phase coherent correction to the
polarizability of isolated systems has been recently theoretically
investigated. Efetov found that it is possible to relate self
consistently this correction to the flux dependence of the screened
potential \cite{Efetov}. Noat {\it et al.} calculated this effect in
the diffusive regime \cite{NRB} and found no flux correction in the
canonical ensemble. In the grand canonical (GC) ensemble no effect is
predicted if the RF pulsation $\omega$ is much smaller than the
inverse relaxation time $\gamma$. However when $\omega \gg \gamma$ the
correction to the polarizability is positive. In particular for
quasi-1D rings, one has \cite{Epsilon}:
\begin{equation} \label{da}
  \frac{\delta_\Phi \alpha^{'}}{\alpha_{1D}}=\frac{\epsilon}{16 \pi^2
  \ln(R/W)} \frac{\Delta}{E_c} \frac{\lambda_s}{W}
\end{equation}
Using supersymmetry techniques Blanter and Mirlin essentially
confirmed these results \cite{Blanter}. Since the rings in our
experiment are completely isolated, the results concerning the
canonical ensemble, in which the effect is predicted to be zero or
very small, should apply. However our experiment shows unambiguously a
decrease of the resonance frequency as we increase the magnetic field
at low field : this corresponds to a {\it positive}
magneto-polarizability. The GC result (eq. \ref{da}) leads
to $\delta_\Phi \alpha^{'}/\alpha_{1D}=0.8 \, 10^{-3}$, which is close
to the experimental value. Therefore the value and the sign of the
effect are consistent with rings considered in the GC case in the
limit $\gamma \ll \omega$. This discrepancy can be related to an
ensemble averaging intermediate between canonical and GC, a situation
called GC-canonical by Kamenev and Gefen \cite{Kamenev} : if the
system is brought to a GC equilibrium at a certain value of the
magnetic flux and then submitted to a time dependent flux whose time
scale is faster than the particle equilibration time the response of
the system can be identical to the GC case in the limit $\gamma \ll
\omega$.  Another possibility is that the mathematical cancellation
responsible for the absence of magneto-polarizability in the canonical
ensemble disappears when one does not have the condition $\omega \ll
\Delta$.
\figtrois

We find that the polarizability is greater in a magnetic field than in
zero field.  This can be related to the flux dependence of screening
within a simple model in which the Thomas-Fermi screening length
is flux dependent. As the correction to the polarizability due to
screening is negative and of the order of $\lambda_s/a$, an increase
of polarizability corresponds to a decrease of the screening length.
Hence the charges are more concentrated on the edges of the sample in
the presence of a magnetic flux. This phenomenon has to be related to
the disappearance of weak localization and thus enhancement of the
metallic character of the sample in the presence of a magnetic field.

Concerning the quality factor our result are, at least for the small
electronic density, in agreement with what is predicted for the
dissipative part of the magneto-polarizability \cite{Noat}. For low
magnetic field we observe a negative $\delta_\Phi(1/Q)$ which
corresponds to a negative $\delta_\Phi \alpha^{''}$ according to (2).
The measured ratio $\delta_\Phi \alpha^{''}/\delta_\Phi \alpha^{'}=
-0.29$ for the illumination 0. Note that this ratio does not depend on
the electric coupling coefficient and hence can be determined with a
good accuracy. When $\Delta$ is less than the temperature, $\delta_\Phi
\alpha^{''} / \delta_\Phi \alpha^{'}$ is related to the level spacing
distributions function \cite{Sivan,Noat} which obeys universal rules
of random matrix theory \cite{Gorkov,Mehta}~:
\begin{equation} \label{RMT}
\frac{\delta_\Phi \alpha^{''}}{\delta_\Phi \alpha^{'}}=\frac{2 \pi
  \omega}{\Delta}(R^{GUE}(\frac{\pi \omega}{\Delta})-R^{GOE}(\frac{\pi
  \omega}{\Delta}))
\end{equation}
$R^{GUE}$ is the two level correlation function in the gaussian
unitary ensemble and $R^{GOE}$ in the gaussian orthogonal ensemble.
This formula is valid in the limit $\gamma \ll \omega$ and $\gamma \ll
\Delta$ and yields to $\delta_\Phi \alpha^{''}/\delta_\Phi \alpha^{'}
= -0.26$ which is close to the experimental value. This is a good
indication that we are effectively in the regime $\gamma \ll \omega$
and $\gamma \ll \Delta$ as confirmed by temperature dependence of the
signal.
\figquatre

The first harmonic of $\delta_\Phi f/f$ obeys an exponential decay
with a typical temperature scale of 90 mK, independent of illumination
(Fig \ref{fig:dftemp}).  Taking the temperature dependence of the
first harmonic to be $\exp{(-2L/L_\Phi)}$, like in weak localization
\cite{Aronov}, and supposing that all the temperature dependence comes
from $L_\Phi$, one has $L_\Phi \propto 1/T$. We deduced
$\gamma=1/\tau_\Phi=D/L_\Phi^2 \simeq 0.8$ mK at 18 mK. It increases
like $T^2$, in agreement with theoretical predictions on the
broadening of single electron energy levels due to electron-electron
interaction in a quantum dot\cite{Sivan2}, and remains below $\Delta$
up to $T=180$ mK. At 50 mK, $L_\Phi \simeq 18\,\mu$m which is larger
than the coherence length deduced from weak-localization measurements
on connected samples($L_\Phi \simeq 6.5\,\mu$m at the same
temperature). In this latter case one has a 1D geometry and a
broadening of the energy levels due to the coupling with reservoirs
whereas in an isolated ring the energy spectrum is discrete.

Illumination increases the electronic density in the rings.  We start
from a situation where the rings are empty or in an electronic
localized state. The increase of the signal observed at low
illumination corresponds to the repopulation of these depleted
rings. The subsequent decay of the signal can be attributed to the
increase of the average conductance of the rings and the $1/g$
dependence of the magneto-polarizability in the diffusive regime ({\it
cf} formula 4. $g$ is the dimensionless conductance, defined as
$g=E_c/\Delta$).  

It is instructive to compare these results with previous measurements
\cite{Noat,Reulet95} on similar array of rings coupled to a multi-mode
strip line resonator sensitive to both electric and magnetic
response. Similar amplitudes of the flux dependence of the resonance
frequency are found in both experiments indicating that the
electric part of the response of the rings is at least of the same
order of magnitude as the magnetic one. We plan to measure this latter
quantity  by coupling the rings to the inductive part of
the resonator.

To conclude we have measured the flux dependent part of the
ac-polarizability of mesoscopic rings down to 18 mK. Both the
non-dissipative and dissipative parts of the polarizability exhibit a
small correction periodic in flux with a period of half a flux
quantum in a ring. The correction of the non-dissipative part is
positive in low magnetic field in agreement with theoretical
predictions in the GC ensemble in the limit $\gamma \ll
\omega$. It indicates a better screening of the electric field in the
presence of magnetic flux. The correction to the dissipative part is
negative for low field, at least for low electronic density. The
effect on the polarizability is qualitatively consistent with a $1/g$
dependence over the dimensionless conductance $g$. These corrections
are sensitive to temperature, with a typical scale of 90 mK. It
would be interesting to pursue these studies in the low frequency
regime.

We thank B. Etienne for the fabrication of the heterojunction and
acknowledge fruitful discussions with M. Nardone, L.P. L\'evy,
L. Malih, A. Mac Farlane, S. Gu\'eron and L. Limot.


\begin{thebibliography}{}
\bibitem{testRPE} This separation between magnetic and electric
  response has been checked by deposition of a paramagnetic system
  (DPPH) alternatively on the capacitive and inductive part of the
  resonator. A magnetic spin resonance signal was only observed when
  DPPH was on the inductive part.
\bibitem{Epsilon} The relative dielectric constant of the substrate is
  a relevant quantity for the flux correction to the polarizability
  because the screening strongly depends on the material in which it
  takes place (in our case GaAs).
\end{thebibliography}
\end{document}